\documentclass[doublespacing]{elsart}
\usepackage{epsfig}
\usepackage{amssymb}
\usepackage{amsfonts}
\usepackage{subfigure}
\usepackage{float}
\usepackage{latexsym,hyperref}
\begin{document}

\runauthor{A.\ Barresi et al.}
\begin{frontmatter}
\title{A finite temperature investigation of dual superconductivity
in the modified SO(3) lattice gauge theory}
\author[pi]{A.~Barresi},
\author[be]{G.~Burgio},
\author[ge]{M.~D'Elia} and
\author[be]{M.~M\"uller-Preussker}

\address[pi]{Dipartimento di Fisica dell'Universit\`a di Pisa and INFN, Pisa, 
Italy}

\address[ge]{Dipartimento di Fisica dell'Universit\`a di Genova and INFN, 
Genova, Italy}

\address[be]{Humboldt-Universit\"at zu Berlin, Institut f\"ur Physik, 
Berlin, Germany}

\begin{abstract}
We study the $SO(3)$ lattice gauge 
theory in 3+1 dimensions with the adjoint Wilson action modified by a 
$\mathbb{Z}_2$ monopole suppression term and by means of the Pisa 
disorder operator. We find evidence for a finite temperature deconfinement
transition 
driven by the condensation of $U(1)$ magnetic charges. A finite-size 
scaling test shows consistency with the critical exponents
of the 3D Ising model.
\end{abstract}
\end{frontmatter}

Lattice $SU(N)$ pure gauge theories in the fundamental
representation undergo a finite temperature
deconfinement phase transition \cite{McLerran:1981pk,Kuti:1981gh} signalled 
by the spontaneous breaking of the global center symmetry 
$\mathbb{Z}_N$ \cite{Polyakov:1978vu,Susskind:1979up}. 
It is an interesting question 
whether this happens also for the theory in the 
adjoint representation $SU(N)/\mathbb{Z}_N$ \cite{Smilga:1994vb}.
In the latter case the center is trivial and naively there is no 
global symmetry to be broken. Moreover, although both the fundamental 
and the adjoint $SU(N)$ lattice theories have the same 
naive continuum limit,
non-perturbative investigations carried out for Wilson 
\cite{Bhanot:1981eb,Greensite:1981hw} as well as for Villain 
discretizations \cite{Halliday:1981te} 
showed a different behavior. For the adjoint cases no evidence
for a finite temperature phase transition was found, whereas 
a bulk transition separating the strong from the weak 
coupling regions appeared for both kinds of discretizations. 

The bulk transition was explained in terms of a condensation
of lattice artifacts - $\mathbb{Z}_2$ monopoles \cite{Halliday:1981te}.
It was argued that the finite temperature transition could be 
overshadowed by the bulk one and $\mathbb{Z}_2$ 
monopole suppression terms were thus proposed \cite{Halliday:1981tm}.
More recently the Villain mixed fundamental-adjoint SU(2) model with a
monopole (and vortex) suppression has been reinvestigated 
\cite{Datta:1998nv,Datta:1999np,Datta:1999ep} 
and first evidence for a deconfinement transition in the Ising 3D 
universality class, at least for strong coupling ($N_\tau=2$), 
was given. Moreover, in the weak coupling region negative states 
of the Polyakov loop in the adjoint representation were found 
\cite{Datta:1998nv,Cheluvaraja:1996zn} and later linked to the non-trivial 
twist sectors of the theory \cite{deForcrand:2002vs}, 
given that on the hypertorus $T^4$ the partition function of the $SO(3)$ 
theory in the Villain formulation with complete $\mathbb{Z}_2$ monopole 
suppression was shown to be equivalent to that of the $SU(2)$ theory 
in the fundamental representation when summed over all twist sectors
\cite{Mack:1982gb,Tomboulis:1981vt,Kovacs:1998xm,Alexandru:2000wp}. 

\begin{figure}[htb]
\begin{center}
\includegraphics[angle=0,width=0.6\textwidth]{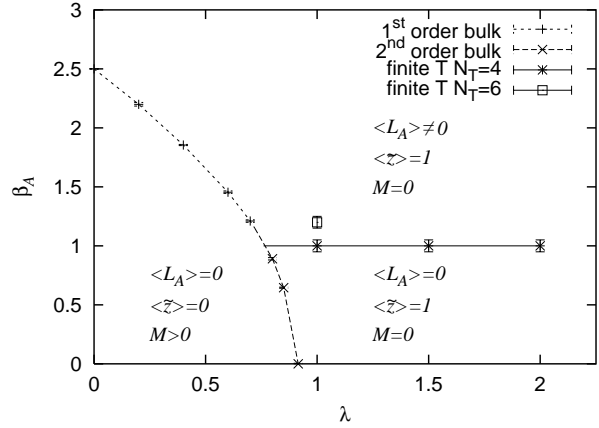}
\end{center}
\caption{Phase diagram of the modified $SO(3)$ Wilson theory with 
$\mathbb{Z}_2$ monopole suppression as seen in the trivial twist sector. 
The phases are characterized by (non)vanishing values of 
the monopole density $M$, the average adjoint Polyakov loop 
$\langle L_A \rangle$ and the averaged electric twist variable 
$\langle \tilde{z} \rangle$ (see \cite{Barresi:2003jq}).}
\vspace*{0.5cm}
\label{diagram}
\end{figure}

In a recent paper we have reinvestigated the $SO(3)$ theory using
the Wilson action and a $\mathbb{Z}_2$ monopole suppression term with a 
``chemical potential'' $\lambda$ 
\cite{Barresi:2001dt,Barresi:2002un,Barresi:2003jq,Barresi:2003yb}. 
The phase diagram in the
$\beta_A-\lambda$ plane was studied at zero and finite temperature 
monitoring the tunneling between twist sectors and its disappearance for
strong enough $\mathbb{Z}_2$ monopole suppression. For $\lambda \ge .85$
we have found a strong indication for the existence of a finite temperature 
deconfinement transition although having restricted the simulation to a
{\it fixed} (e.g. trivial) twist sector. 
The proposed phase diagram is redrawn in Fig. \ref{diagram}.
Since no proper order parameter was available a determination 
of critical exponents was not intended, and there was no answer given to 
the question about the underlying confinement mechanism.

A disorder parameter related to Abelian monopole condensation 
in the dual superconductivity picture of confinement 
\cite{Nambu:1974zg,Mandelstam:1975hb,'tHooft:1975th} has been
devised by the Pisa group some time ago 
\cite{DiGiacomo:1997sm,DiGiacomo:1999fa,DiGiacomo:1999fb,Carmona:2001ja,Carmona:2002ty}. 
It is the 
vacuum expectation value of a magnetically charged operator 
$~\langle \mu \rangle~$ shown to be different from zero in the 
confined phase, thus signaling dual superconductivity,
and going to zero at the deconfining phase transition.
Similar parameters have been constructed more recently by Fr\"ohlich and 
Marchetti \cite{Frohlich:2000zp} as well as in the framework of the lattice 
Schr\"odinger functional \cite{Cea:2000zr,Cea:2004ux} 
leading to analogous results. 
The main advantage of these parameters is that they can be applied also 
to full QCD, where the center symmetry is explicitly broken by the fermionic
degrees of freedom and - as we shall show - to the adjoint pure 
gauge theory, where center symmetry becomes trivial. 
In this letter we will use the Pisa disorder operator in order to answer the 
questions raised above for the $SO(3)$ lattice gauge theory.  
In particular we will 
check the critical exponents and whether the dual superconductor scenario 
applies also to this case.


We will study the $SU(2)$ adjoint representation Wilson action
modified by a $\mathbb{Z}_2$ monopole suppression term
\begin{eqnarray} 
S=\frac{4}{3}\beta_{A} \sum_{P} 
  \left(1-\frac{\mathrm{Tr}_{F}^{2}U_{P}}{4}\right)
  +\lambda \sum_{c}(1-\sigma_{c})\,,
\label{ouraction}
\end{eqnarray}
where the product
$\sigma_{c}=\prod_{P\in\partial c}\mathrm{sign}(\mathrm{Tr}_{F}U_{P})$,
taken around elementary 3-cubes $c$, defines the 
$\mathbb{Z}_2$ magnetic charges. Their density can be introduced as 
$M = 1-\langle\frac{1}{N_c}\sum_{c}\sigma_{c}\rangle$
normalized such that it tends to one in the strong coupling region
and to zero in the weak coupling limit, $N_c$ denoting the total 
number of elementary 3-cubes. Although $\sigma_{c}$ is constructed in terms of fundamental 
representation plaquettes, it is a {\it natural} $SO(3)$ 
quantity ensuring that the action (\ref{ouraction}) is center-blind
in the entire $\beta_A-\lambda$ plane.  The link variables
can be represented both by $SO(3)$ or $SU(2)$ matrices, exploiting the property
$\mathrm{Tr}_{A}= \mathrm{Tr}_{F}^2-1$ for the Wilson term or picking
a random $SU(2)$ representative of the $SO(3)$ link to construct 
the $\mathbb{Z}_2$ monopole contribution. 
A standard Metropolis algorithm has been used to update the links.
In \cite{Barresi:2003jq} we have
found a strong indication in favor of finite temperature 
phase transitions with lines moving up with the time-like lattice extent
$N_\tau$ and running away from the bulk transition approximately parallel 
to the $\lambda$ axis (see Fig. \ref{diagram}). 


The Pisa disorder operator 
\cite{DiGiacomo:1997sm,DiGiacomo:1999fa,DiGiacomo:1999fb,Carmona:2001ja,Carmona:2002ty} 
was shown to be a 
reliable order parameter for $SU(2)$ and $SU(3)$ gauge theories in the 
fundamental representation, with and without dynamical quarks,
giving critical exponents in agreement with other order parameters.
Moreover, it can give important informations about the mechanism
which confines quarks into hadrons.
The Pisa disorder operator is motivated by the dual superconductor 
scenario for the QCD vacuum 
\cite{Nambu:1974zg,Mandelstam:1975hb,'tHooft:1975th}, 
driven by the condensation of 
$U(1)$ magnetic charges. The construction of the operator in the 
case of the modified $SO(3)$ theory follows the same line as in the 
fundamental case, so we will avoid going into the details and we will 
refer to the original papers for further details 
\cite{DiGiacomo:1997sm,DiGiacomo:1999fa,DiGiacomo:1999fb,Carmona:2001ja,Carmona:2002ty}.

The idea is to construct a magnetically charged operator $\mu$ which
shifts the quantum field at a given time slice by a classical 
external field corresponding to a magnetic monopole.
The $U(1)$ subgroup of the gauge group which defines the magnetic charge
is selected by an Abelian projection, usually fixed by diagonalizing 
an operator $X$ in the adjoint representation. 
The disorder parameter is defined as
\begin{eqnarray} 
\langle \mu(t)\rangle = \frac{\int (DU)_M e^{-S_M(t)}}{\int (DU) e^{-S}}\,,
\label{pisa}
\end{eqnarray}
where $S_M(t)$ denotes the Wilson action with the space-time plaquettes 
$U_{i4}(\vec{x},t)$ at a fixed time-slice $t$ modified by an insertion of 
an external monopole field 
\begin{eqnarray}
\widetilde{U}_{i4}(\vec{x},t)=U_i(\vec{x},t)\Phi_i(\vec{x}+\hat{i},\vec{y})
U_4(\vec{x}+\hat{i},t) U_i^{\dagger}(\vec{x},t+1)
U_4^{\dagger}(\vec{x},t))\,, 
\label{plaq_shift}
\end{eqnarray}
where
$\Phi_i(\vec{x},\vec{y})=\Omega e^{i T_a b^a_i(\vec{x}-\hat{i},\vec{y})}
\Omega^\dagger $, with
$\Omega$ the gauge transformation which diagonalizes an operator $X$ 
in the adjoint representation. $T_a$ denote the generators 
of the Cartan subalgebra
and $\vec{b}$ the discretized transverse field generated at
the lattice spatial point $\vec{x}$ by a magnetic monopole
sitting at $\vec{y}$. 
We decided to work with the completely random Abelian
projection (RAP) \cite{Carmona:2001ja,Carmona:2002ty} 
in which we do not diagonalize any operator $X$:
it can be thought as a kind of averaging over a continuous
infinity of Abelian projections. 
It should be stressed that only the plaquette contribution to the 
action (\ref{ouraction}) is modified by the insertion of the monopole 
field and not the chemical potential term.
From the definition of $\mu$, making use of an iterated change of variables,
it can be shown that the 
correlation function
$D(\Delta t)=\langle
\bar{\mu}(\vec{y},t+\Delta t)\mu(\vec{y},t)\rangle $
describes the creation of a monopole at $(\vec{y},t)$
and its propagation from $t$ to $t+\Delta t$ 
\cite{DiGiacomo:1997sm,DiGiacomo:1999fa,DiGiacomo:1999fb,Carmona:2001ja,Carmona:2002ty}. 
At large $\Delta t$, by cluster
property, $D(\Delta t)\simeq A\exp(-M \Delta t)+\langle\mu\rangle^2 $.
A non-vanishing $\langle\mu\rangle$ indicates spontaneous breaking
of the $U(1)$ magnetic symmetry and hence dual superconductivity.
In the thermodynamical limit one expects $\langle\mu\rangle\neq 0$ for
$T<T_c$, while $\langle\mu\rangle = 0$ for $T>T_c$ 
if the deconfining phase transition is associated with a transition
from a dual superconductor to a trivial vacuum.

At finite temperature there is no way to put
a monopole and an antimonopole at large distance along the $t$-axis 
as it is done at $T=0$, since at $T\sim T_c$ the temporal 
extent $N_\tau a$ is comparable to the correlation length. Therefore, 
one computes $\langle \mu \rangle$ but with $C^*$-periodic 
boundary conditions in time direction imposed to the numerator in 
Eq. (\ref{pisa}) in order to ensure magnetic charge conservation:
$U_i(\vec{x},N_\tau)=U_i^*(\vec{x},0)$,
where $U_i^*$ is the complex conjugate of $U_i$.
These boundary conditions have been indicated in Eq. (\ref{pisa})
by the index $M$ at the integration measure. They change the sign of the 
term proportional to $\sigma_3$ in the links, creating a dislocation with 
magnetic charge -1 at the boundary which annihilates the positive 
magnetic charge created by the operator $~\mu~$. As a consequence the 
magnetic charge is conserved and everything is consistent.
Of course, the denominator in Eq. (\ref{pisa})  is computed  using
standard periodic boundary conditions.
We used links in the fundamental representation, so we implemented
exactly the above condition, but an 
analogous condition holds also for link variables defined in the 
adjoint representation, i.e.
$U_i(\vec{x},N_\tau)=
(\mathbb{I}_3+2T^2_2)U_i(\vec{x},0)(\mathbb{I}_3+2T^2_2)$;
charge conjugation is realized in both representations
through rotations by an angle $\pi$ around the color 2-axis.

Since $\langle\mu\rangle$  is the average of the exponential of a sum over the 
physical volume, it is affected by huge fluctuations which make it 
difficult to be measured in Monte Carlo simulations.
A way out is to compute the derivative
with respect to the coupling parameter $\beta \equiv \beta_A$, 
which contains all the relevant information
\begin{equation} 
\rho=\frac{d}{d\beta}\log\langle\mu\rangle\;.
\label{pisa2}
\end{equation}
It is given by the difference between the Wilson plaquette action
term $\langle\Pi\rangle$ averaged with the usual measure and the 
modified plaquette action term $\langle\Pi_M\rangle$ averaged with 
the modified measure $(DU)_M e^{-S_M} / \int (DU)_M e^{-S_M}$.
The order parameter can be reconstructed from
\begin{equation}
\langle\mu\rangle =\exp\left(\int_0^\beta \rho(\beta')d\beta'\right)\,.
\label{rhomu}
\end{equation}
Eq. (\ref{rhomu}) tells us that in order to have $\langle \mu \rangle \neq 0$ 
in a confined phase, where the dual magnetic symmetry becomes broken, 
$\rho$ should stay finite for $\beta < \beta_c$ in the thermodynamical 
limit (i.e. in the limit of spatial lattice extent $N_s \to \infty$), 
while a sharp negative 
peak for $\rho$ occuring at $\beta_c$ and diverging for $N_s \to \infty$ 
should signal the phase transition associated 
with the restoration of the dual magnetic symmetry. Above the transition a 
sufficient condition for $\langle \mu \rangle$ to vanish would be to have 
$\rho \to - \infty$ in the thermodynamical limit. In 
\cite{DiGiacomo:1997sm,DiGiacomo:1999fa,DiGiacomo:1999fb,Carmona:2001ja,Carmona:2002ty} for 
the $SU(2)$ case it was argued using perturbation theory that $\rho$ for 
$\beta \to \infty$ reaches negative plateau values linearly scaling with 
$N_s$. 
In ~\cite{D'Elia:2003xn} a more detailed numerical analysis has been
performed for both SU(2) and SU(3) pure gauge theories, showing that,
in the case where $\mu$ is magnetically charged,
$\rho$ diverges (to negative values) linearly in the weak coupling limit, and
more and more rapidly as $T \to T_c$ from above, where it diverges as
$N_s^{1/\nu}$, thus proving that $\langle \mu \rangle $ is strictly
zero for every temperature $T > T_c$, as follows from Eq.~(\ref{rhomu}).
Therefore, when simulating the theory at accessibly large $\beta$ values and 
lattice sizes, a good criterion for $\langle \mu \rangle$ being
exactly zero above the transition temperature, is that $\rho$ keeps 
diverging at least linearly with $N_s$ in a wide range of $\beta$ 
values above the transition.

\begin{figure}[htb]
\begin{center}
\subfigure[$\lambda=0.0$]{
\includegraphics[angle=0,width=0.48\textwidth]{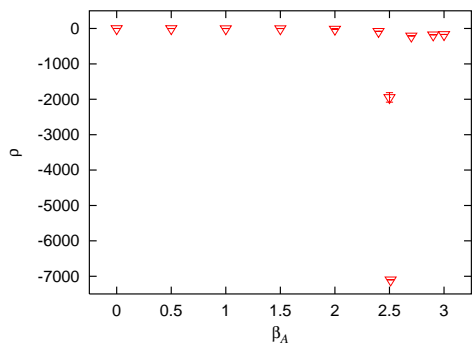}}
\subfigure[$\lambda=0.4$]{
\includegraphics[angle=0,width=0.48\textwidth]{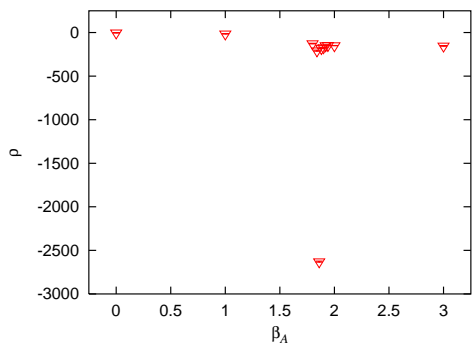}}
\subfigure[$\lambda=0.7$]{
\includegraphics[angle=0,width=0.48\textwidth]{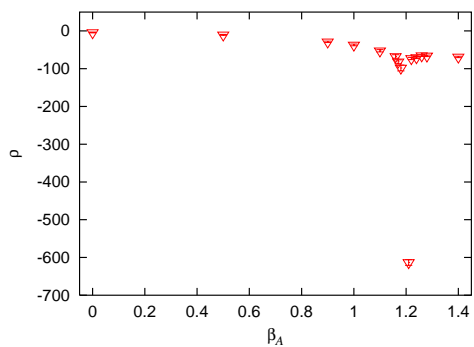}}
\subfigure[$\lambda=0.8$]{
\includegraphics[angle=0,width=0.48\textwidth]{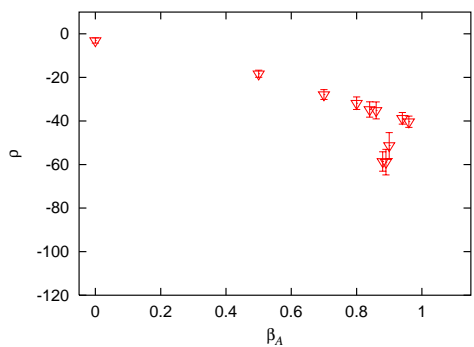}}
\end{center}
\caption{Disorder operator $\rho$ computed at different values of the 
chemical potential $\lambda$ at finite temperature in the asymmetric 
volume $V=4\times12^3$.} 
\vspace*{0.5cm}
\label{lambda}
\end{figure}
We will now present our numerical results. First of all
we studied the bulk transition (compare with Fig. \ref{diagram})
for some $\lambda$-values between $0$ (no monopole suppression)
and $0.8$ (partial monopole suppression) by varying $\beta_A$. 
We do this for finite temperature ($N_\tau=4$) to investigate
the interplay between the bulk transition and the finite temperature
one. Fig. \ref{lambda} shows 
a clear dip at the location of the bulk transition, 
in agreement with what was found in previous works with other
observables. Although the dip in $\rho$ decreases in magnitude
with increasing $\lambda$, i.e. as we are suppressing more and
more the $\mathbb{Z}_2$ monopoles, the lattice artifacts
are still present and overshadow the finite temperature transition,
assuming the latter exists. Therefore, one cannot yet see neither the 
scaling of the physical transition with the temperature 
nor the finite-size scaling of the operator with the spatial volume, 
although one expects that on large enough
lattices the two transitions should decouple also for
these values of the couplings \cite{Barresi:2003jq}. Anyway we can conclude
that a condensation of $U(1)$ magnetic charges takes place 
below the bulk transition line.

\begin{figure}[thb]
\begin{center}
\subfigure[]{
\includegraphics[angle=0,width=0.48\textwidth]{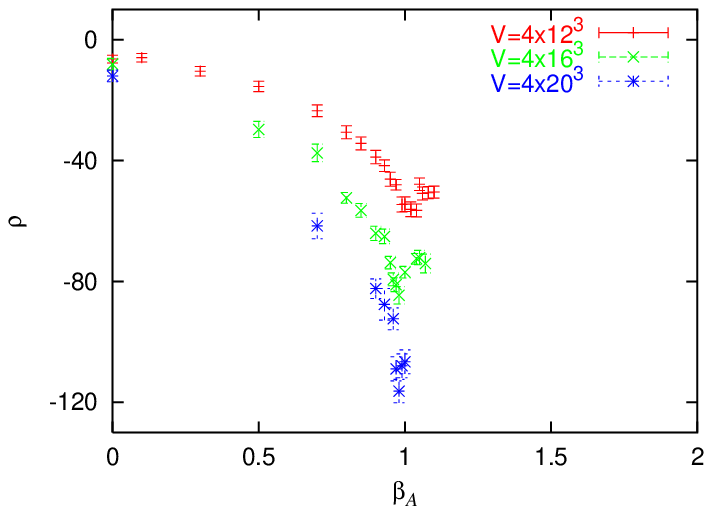}}
\subfigure[]{
\includegraphics[angle=0,width=0.48\textwidth]{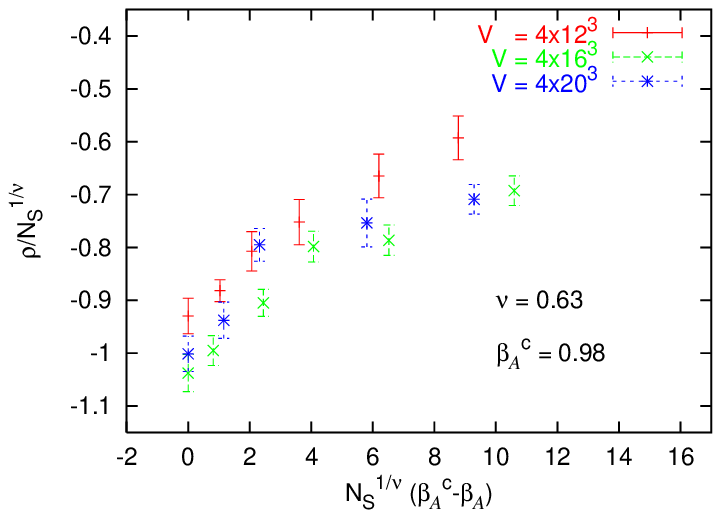}}
\subfigure[]{
\includegraphics[angle=0,width=0.48\textwidth]{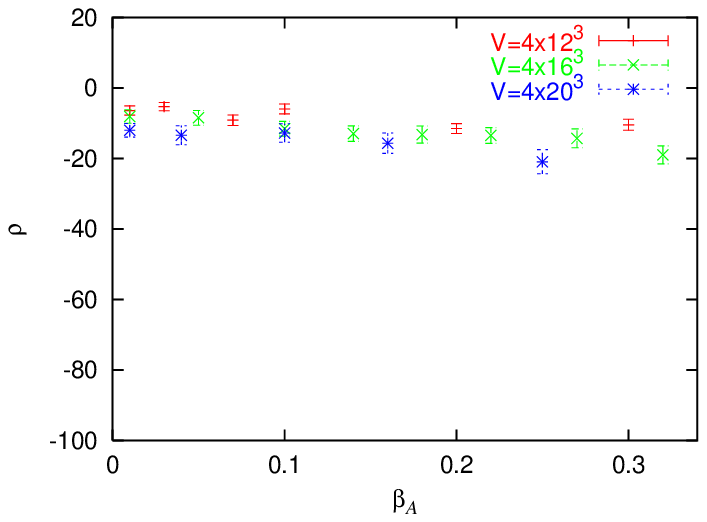}}
\subfigure[]{
\includegraphics[angle=0,width=0.48\textwidth]{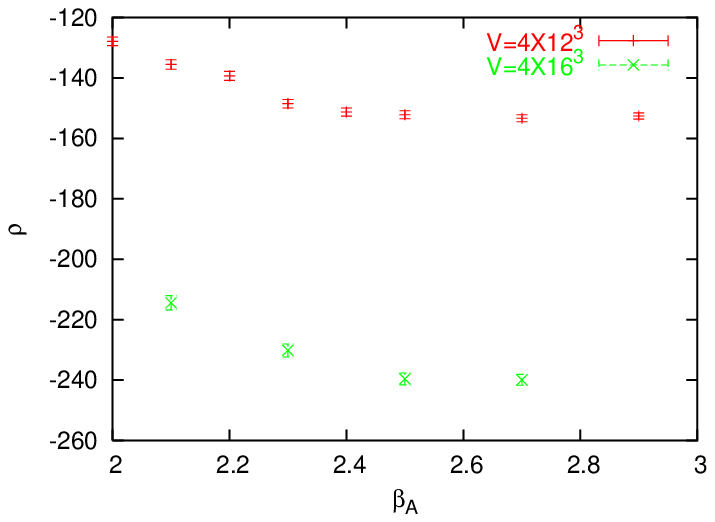}}
\end{center}
\caption{$\rho$ computed in the trivial twist sector
at finite temperature ($N_\tau=4$) for different values of the
spatial volume and for chemical potential $\lambda=1.0$ (a).
Finite-size scaling analysis for $\rho$ (b). The strong and 
weak coupling regions are highlighted in (c) and (d) respectively.} 
\vspace*{0.5cm}
\label{nt4}
\end{figure}

Let us now turn to the more interesting case of stronger monopole 
suppression ($\lambda=1.0$), where the decoupling of the bulk from the finite 
temperature transition occurs already at reasonable volumes.
Again we vary the temperature through $\beta_A$.
We have kept the system fixed in the trivial twist sector so to
compare with the results for the adjoint Polyakov loop 
and the distribution of the fundamental Polyakov loop variable in 
\cite{Barresi:2003jq}. 
As one can see from Fig. \ref{nt4} (a), at fixed $N_\tau=4$
the parameter $\rho$ shows a dip around $\beta_A \simeq 1.0$ which 
becomes deeper and deeper with increasing spatial volume. $~\rho~$ stays
finite in the low $\beta_A$ region as the spatial volume is increased,
as can be inferred from Fig. \ref{nt4} (c), while the data for high $\beta_A$ 
are consistent with its saturation at negative plateau values
diverging more than linearly with $N_s$, as Fig. \ref{nt4} (d) shows. 
This gives us clear evidence that
also in this case, like for the lattice theory in the fundamental 
representation, condensation of $U(1)$ magnetic charge takes place
below the deconfinement phase transition and disappears above $T_c$.  
The vacuum is a dual superconductor in the confined phase and becomes trivial
in the deconfined one. It shows that the Pisa disorder operator is a 
meaningful order parameter also in a center-blind theory and for fixed twist. 
Moreover, the position of the dip is well-consistent with the results 
found in our previous work \cite{Barresi:2003jq}.
\begin{figure}[thb]
\begin{center}
\subfigure[]{
\includegraphics[angle=0,width=0.48\textwidth]{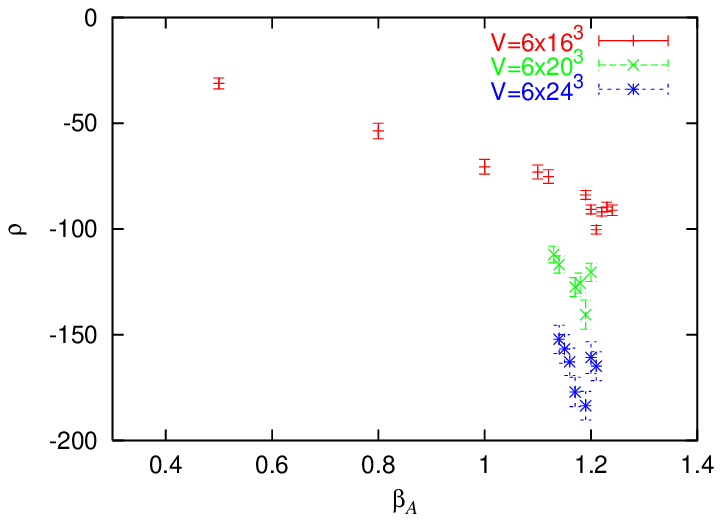}}
\subfigure[]{
\includegraphics[angle=0,width=0.48\textwidth]{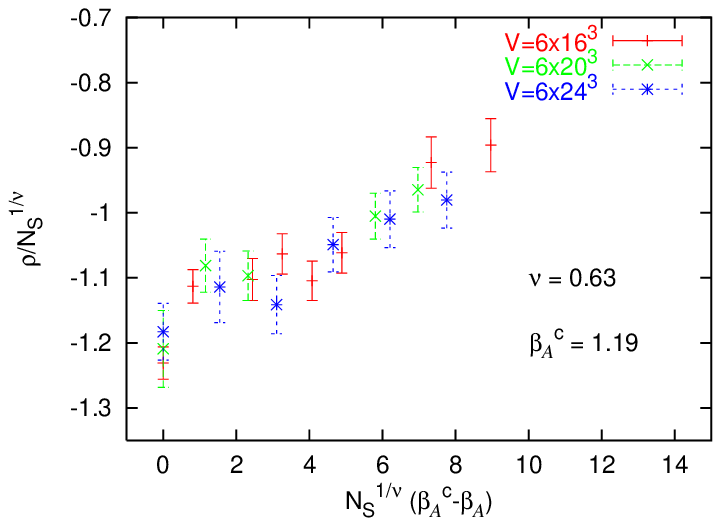}}
\end{center}
\caption{$\rho$ computed in the trivial twist sector
at finite temperature ($N_\tau=6$) for different values of the
spatial volume and chemical potential $\lambda=1.0$ (a). 
Finite-size scaling analysis for $\rho$ (b).}
\vspace*{0.5cm}
\label{nt6}
\end{figure}
The statistics is not sufficient in order to determine
the critical exponents independently, 
but as one can see from Fig. \ref{nt4} (b),
using the known critical exponent for the 3D Ising model, 
$\nu=0.63$ and estimating the critical adjoint coupling
$\beta_A^c=0.98$ the data show a reasonable 
finite-size scaling behavior with the spatial volume.
We have also checked the scaling with respect to the continuum limit 
for varying $N_\tau$, making some simulations by fixing
$N_\tau=6$ and $N_s=16,20,24$.
As one can see from Fig. \ref{nt6} (a) the dip occurs at a larger 
$\beta_A$ value than for $N_\tau=4$ and becomes again deeper by
increasing $N_s$, in agreement with a finite 
temperature phase transition. In this case we estimate
$\beta_A^c=1.19$. Fig. \ref{nt6} (b) shows the quality of finite-size 
scaling assuming the Ising model value for the critical index. 
Our results are in agreement with the finite-size scaling observed 
for the specific heat in the Villain action case for $N_\tau=2$
\cite{Datta:1998nv,Datta:1999np,Datta:1999ep}.

We can conclude that our investigation of the modified 
lattice $SO(3)$ gauge theory with Wilson action and  
$\mathbb{Z}_2$ monopole suppression
using the Pisa disorder operator has confirmed our previous results
\cite{Barresi:2003jq}. Although having restricted to the fixed trivial twist
case we find a clear indication for the existence of a finite
temperature transition decoupled from the notorious bulk transition.  
The critical behavior at such a transition reasonably agrees with 
the critical exponents of the 3D Ising model. The nature of the
Pisa disorder parameter we used tells us that the transition is related to
a condensation of $U(1)$ magnetic charges for $T<T_c$ which disappears above 
the transition, proving that 
the dual superconductor scenario is a good model of confinement 
also for the adjoint theory. Of course, a final answer can be given
only after all twist sectors will be taken into account simultaneously.
This study is currently under way. Moreover, it would certainly be interesting
to compute also the free energy of an (extended) center vortex in order
to check how the vortex condensation mechanism works in this case.

We thank A. Di Giacomo, L. Del Debbio, and P. de Forcrand 
for valuable comments and discussions.

\vspace{-0.5cm}

\bibliographystyle{amsplain}
\bibliography{dj_rev}
\end{document}